\documentclass{article}

\usepackage[utf8]{inputenc}

\usepackage{arxiv}

\usepackage{amsmath}
\usepackage{hyperref}
\usepackage{url}
\usepackage{graphicx}
\usepackage{xcolor}
\usepackage{booktabs}
\usepackage{lipsum}
\usepackage{enumitem}
\usepackage{float}
\usepackage{array}
\usepackage{tablefootnote}
\usepackage{comment}
\usepackage{hyperref}
\usepackage{amsfonts}       
\usepackage{nicefrac}       
\usepackage{microtype}      

\usepackage{natbib}
\usepackage{doi}

\usepackage{xurl}

\definecolor{fig3pink}{RGB}{255, 66, 161}
\definecolor{fig3green}{RGB}{0, 108, 101}
\definecolor{fig3blue}{RGB}{117, 177, 212} 
\definecolor{fig3purple}{RGB}{128, 12, 70}

\title{Making Chant Computing Easy: \\ CantusCorpus v1.0 and the PyCantus Library}
\author{%
Anna Dvořáková \\ Charles University \\ Prague, Czech Republic \And %
~Tim Eipert \\ Julius-Maximilians-Universität Würzburg \\ Würzburg, Germany \And %
~Debra Lacoste \\ Dalhousie University \\ Halifax, Nova Scotia, Canada \And %
~Jan Hajič jr.* \\ Charles University \\ Prague, Czech Republic \\ \texttt{hajicj@ufal.mff.cuni.cz}
}

\date{}

\begin{document}


\twocolumn[{%
\maketitle
\begin{abstract} 
Digital Gregorian chant scholarship has for decades enjoyed the privilege of a large digital resource cataloguing chant sources: the Cantus ecosystem,
with nearly 900,000 chants catalogued across more than 2000 sources.
The Cantus Database data model and the Cantus ID mechanism has been adopted by 18 more chant databases, jointly accessible through the Cantus Index interface. 
However, this data has only been available piecemeal via the individual online user interfaces;
computational methods have so far had only a limited opportunity to process these immense resources.
To overcome this hurdle, we compiled CantusCorpus v1.0, a dataset that combines everything that was available across the Cantus Index-centered network of databases as of mid-2025, and we have also provided the code for updating the dataset as the databases grow.
We then created the lightweight PyCantus library for working with this data. 
PyCantus decouples the data model from the Cantus codebase and thus allows integration of further chant data sources, which we illustrate with harmonising pilot data from the Corpus Monodicum project.
Computational chant research is attractive --- and CantusCorpus v1.0 and PyCantus are infrastructures that should make work in this field more transparent, replicable, and accessible to digital humanities practitioners beyond chant scholars themselves.
\end{abstract}
\hspace{24pt}%
}]


\section{Introduction}
\label{sec:introduction}

Gregorian chant,\footnote{Also, Frankish-Roman or Romano-Frankish chant, from the tradition begun in the Carolingian period with the melding of Frankish and Roman liturgies.} the liturgical monody of the Latin church, is one of the best-preserved and largest European traditions, with more than 30,000 extant manuscripts \citep{helsen2014chantOmr} and likely some 500,000 having been produced in the Middle Ages \citep{huglo2009statistics}.

Gregorian chant's standardised traditions provide material that is fairly straightforward to organise as a database.\footnote{To learn more about Gregorian chant tradition and its structure, refer to \cite{hiley2009gregorian}.
}
Its sacred nature within a literate institutional culture and highly formalised integration into liturgy mean that much of the tradition consists of clearly defined units of repertoire --- chants (a pairing of text and melody). Its written record defines relevant categories and adheres to them: most importantly genre (antiphon, responsory, etc.), defined by function and position in liturgy; ``chant slots'' within the liturgical calendar (feast, such as Christmas Day or feasts of individual saints, and the specific liturgies within that feast); and a highly consistent manuscript tradition (which then transitioned into print with the same principles of book organisation).

\begin{figure*}[]
    \centering
    \includegraphics[width=1.0\linewidth]{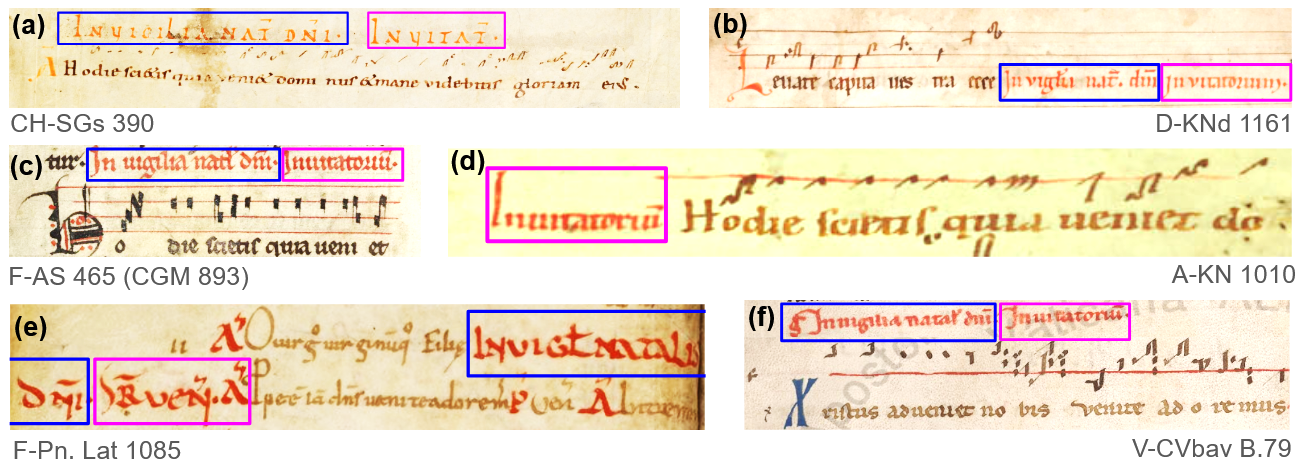}
    \caption{Liturgy defines positions for chant in each day of the year. Different communities of practitioners might, however, use different chants in some of these positions. Each manuscript, therefore, documents how a particular ecclesiastical community placed chants in the liturgy. This assignment is communicated via \textit{rubrics}, instructions in red ink that indicate specifics such as feasts, services during the day, or indications of genre. Comparing chants in the same liturgical position in multiple manuscripts reveals some variety: the Invitatory antiphon (rubric parts highlighted by pink rectangles) for the feast \textit{Vigilia Nativitatis Domini} (rubric parts highlighted in blue), with chant \textit{Hodie scietis} in (a), (c), (d), \textit{Levate capita vestra} in (b), \textit{Prope est jam dominus} in (e), and \textit{Christus adveniet nobis} in (f).}
    \label{fig:simple_hodie}
\end{figure*}

This implies that the vast majority of the written records of Gregorian chant can be unambiguously identified as specific fixed units of the tradition.
Furthermore, each such element is identified \textit{only} by its text, genre, and melody, and in most cases there is just one melody for the given text in a given genre and liturgical position (which can be ascertained from the organisation of a manuscript).
Therefore, one can look at any two chant records in different manuscripts that may be thousands of miles and several centuries apart and still be able to unambiguously say whether these are two records of the same chant, or not.\footnote{With the exception of some later genres --- such as sequences, which are strophic compositions that combine text more flexibly.}
This interplay between individual chants from the Gregorian repertoire, liturgy, and manuscripts that record the use of chants in liturgy is illustrated in \autoref{fig:simple_hodie}.

The internal homogeneity makes it relatively straightforward to catalogue Gregorian chant digitally.
Identifying instances of the same chant in manuscripts has been a regular part of working with chant, notably in recent history, e.g. in the editorial efforts of Solesmes \citep{froger1978critical}, in the \textit{Corpus Antiphonalium Officii} \citep{hesbert1963CAO}, and in the digital domain implemented most notably by the Cantus ID mechanism \cite{lacoste2012cantus}.

The second factor that makes the digital domain ideal to study chant is the scale of this tradition. This is clear on the analytical side, where the volume of data is too large to process manually (downselecting to sets of ``representative'' sources does make larger-scale work like the Graduale Romanum edition possible \citep{froger1978critical}, but it was still a decades-long process). This is also a factor when cataloguing; the number and sizes of extant manuscripts necessarily make this a long-term collaborative effort.
The consistency afforded and enforced by a 
database with online access is ideal to ensure that individual cataloguing efforts can, in fact, be accumulated over time,
and that individual efforts limited by time or funding keep contributing to the field after they end.
Furthermore, the ability to revisit entries is crucial for long-term data quality.

This digital opportunity --- and necessity --- was identified already in the 1980s, and the Cantus Database \citep{lacoste2012cantus} was conceived (see \autoref{sec:cantusecosystem}). 
The implementation of the Cantus Index interface for joint search across all these databases (19, to date) has revolutionised chant research \citep{lacoste2022cantus}. 
Its importance is illustrated by the 61 articles and 8 books published just during the last year that report using Cantus websites.\footnote{According to the unpublished report: Jennifer Bain, Andrew Hankinson, and Debra Lacoste, ``Annual Research Platforms and Portals Progress Report,'' Digital Research Alliance of Canada, 5 Nov 2025.}
There are likely more than 2,000 active users per month on Cantus Index and 8,000 on the Cantus Database.\footnote{From Google Analytics; after subtracting estimated bot activity.}
The Cantus Index federated search hub now reports over 1,150,000 catalogued chants.\footnote{\url{https://cantusindex.org/stats} [2. 2. 2026]}

\subsection{The gap we address}

Despite the abundance of digital chant records online, \textbf{\textit{computational} chant scholarship} (the aforementioned analytical use of the digital domain) \textbf{is unable to take full advantage of this resource,} because the data is accessible only as individual items through user interfaces of individual database websites and Cantus Index, while for computational work, one would need to load the content of the entire network of databases at once.
Furthermore, while chant records themselves are relatively standardised, 
source records are not: 
for instance, information about the provenance and dating of individual manuscripts --- stored at the source level  --- is essential for studying the diversity and development of the Gregorian repertoire, but protocols and data exports for such fields have not yet been codified and normalised across the Cantus Index network.

Computational chant scholarship has seen multiple new works appear recently 
(see \autoref{subsec:relatedwork:tools}), with the potential to take long-standing ideas about chant transmission and melody on more solid empirical ground with data of much larger scopes.
But to realise this potential, \textbf{chant data must first be made available for computational processing.}

\subsection{The Contribution}

To close this gap, we contribute the \textbf{CantusCorpus v1.0 dataset}. It consists of 888,010 chant records and 2,278 source records,
reflecting the state of Cantus Index and its attendant network of databases as of May~20th,~2025. It includes the Extraction-Transformation-Load (ETL) code for building updated versions of the dataset from the Cantus ecosystem.

To make programmatic manipulation of this data simple, consistent, and accessible, we provide the \textbf{PyCantus library}, which implements the data model for CantusCorpus v1.0 (\autoref{sec:thedataset}) and a unified data filtering interface that produces static artefacts to support experiment replicability
and subset reuse across applications developed on top of PyCantus (\autoref{sec:library}).

\begin{figure}[t]
    \centering
    \includegraphics[width=0.99\linewidth]{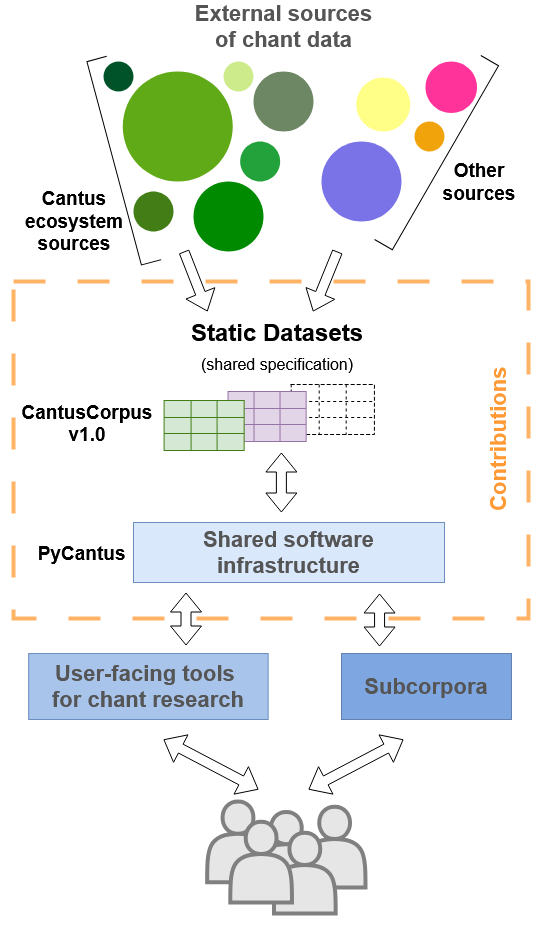}
    \caption{A simplified schema of our contribution.}
    \label{fig:simple_schema}
\end{figure}

The data\footnote{Long-term archival version: \url{http://hdl.handle.net/11234/1-6041}, development: \url{https://github.com/dact-chant/CantusCorpus}} is provided under a Creative Commons 4.0 license, in accordance with Cantus Index licensing, and the PyCantus library is open-source.\footnote{\url{https://github.com/dact-chant/PyCantus}}
While CantusCorpus v1.0 has been collected from the Cantus network, by implementing PyCantus independently on the Cantus Index codebase, we make integration of other chant resources easier, as illustrated by a case study with data from the Corpus Monodicum project (see \autoref{sec:monodicum}).
The structure of our contributions is shown in \autoref{fig:simple_schema}.

\subsection{Why CantusCorpus?}

\noindent What is the value of a dataset that merely compiles a set of existing, related chant databases?

\begin{enumerate}
    \item \textbf{Big questions.} 
    While all the data records are held in the individual sites connected to the Cantus Index hub, they are not \textit{readily} available for computational processing. Even with direct access to the Cantus Index back-end (which obviously cannot be realistically granted), the data are still distributed on individual sites.
    This unnecessarily limits computational chant research in scenarios like ``Does the relationship of tropes to medieval politics \citep{eipert2023tropers} apply to other genres of chant repertoire?'', ``Is the centonization theory of chant melody \citep{ferretti1934estetica, levy1970italian, treitler1975centonate} more likely than not \citep{lanz2025segmentation}?'', or ``How many distinct chants remain to be discovered? \citep{moss2025rest}''. 
    In fact, some results of \cite{moss2025rest} were already revealed as artifacts of limited data compared to CantusCorpus~v1.0 in PyCantus Tutorial 4 (see~\autoref{subsec:library:tutorials}).
    Without having all the data in one ready-to-use unified dataset, research will be hampered by database boundaries --- despite the raw data being in existence.
    
    \item \textbf{Ease of use.} Having a clearly defined dataset can also help
    attract and retain methods-focused collaborators from the field of MIR or data science more broadly.
    
    \item \textbf{Transparency.} The Cantus ecosystem is continuously growing and changing. However, empirical research requires reproducibility.
    Having a dataset unambiguously identified by a DOI or similar mechanism is essential for replicating experiments.
    
    \item \textbf{Standardisation.} Only upon trying to bring together data from across the entire Cantus network do we see that within what is a well-defined data model, there are still significant differences in how databases apply values for individual chant record fields and what information they expose. 
    This work has, as a ``side effect'' of understanding this, the production of a roadmap towards further standardisation across the Cantus ecosystem. 
    The CantusCorpus scraping infrastructure makes it easy to produce new versions of the dataset, expanded and perhaps with more controlled vocabularies as they get instituted across the Cantus network.
    
    \item \textbf{Outreach.} Beyond chant scholarship, the Cantus ecosystem belongs among the largest resources for digital musicology in general, second perhaps only to RISM. Thus, making it easy to start working with this data can also be valuable in educational settings. As digital humanities (DH) programmes proliferate, making chant an attractive DH topic for students with both a humanities background and a technical one would help sustain chant scholarship itself.
\end{enumerate}

\subsection{Why PyCantus?}

Having argued the value of compiling the various database contents into a dataset, we now justify the PyCantus library.

\begin{enumerate}
    \item \textbf{Ease of use.} While there are innumerable ways to work with a CSV file, having a dedicated library that standardises common steps with chant material makes it easier to read and reproduce one's own work as well as the work of other chant scholars.
    
    \item \textbf{Reproducibility.} The PyCantus filtering mechanism ensures that experiments can be run on updated datasets with the same inclusion/exclusion criteria consistently applied for chants and sources, and that preprocessing procedures can be shared as static configuration files. 
    This complements the \textit{Transparency} goal mentioned above.

    \item \textbf{Independence.} We implement the data model of PyCantus without tying it to the Cantus codebase itself. Chant is so thoroughly formalised that different conceptualisations of the tradition may in fact be mutually compatible, and thus the PyCantus data model can be used more broadly.
    To illustrate this, we built a pilot compatible dataset from data of Corpus Monodicum, a project entirely independent of Cantus (see \autoref{sec:monodicum}).
    
    \item \textbf{Ease of development.} Multiple digital chant projects have been ``reinventing wheels'' by implementing near-equivalent chant data models: both the Cantus DB and Cantus Index back-ends, other databases still running on Drupal CMS, and non-database applications such as ChantDigger\footnote{\url{https://github.com/timeipert/chantdigger-restored}}, ChantMapper \cite{dvorakova2025chantmapper} and ChantLab \cite{lanz2025chantlab}, or CEAP \cite{hornby2022CEAP}. The data model and filtering mechanism can be used in \textit{any} Python-based application, while being a lightweight dependency.
\end{enumerate}

\vspace{3mm}

\noindent
The combined value of CantusCorpus~v1.0 and PyCantus for musicological work over existing interfaces becomes clearest for queries combining criteria on both repertoire and sources. E.g., if one wanted to compare major Marian feast repertoire between the 12th and 14th centuries and test for the influence of Cistercians on the repertoire of these feasts, compiling the necessary sub-corpora from the Cantus Index user interface would take hours of copy-pasting (and detours for manuscript metadata to individual database websites). One would then have to do that again for a control group of different feasts. With PyCantus, one would only create filter files with the corresponding inclusion/exclusion criteria and an analysis script --- and share them for anyone else to replicate.

PyCantus is also flexible enough to serve use cases beyond chant scholarship.
A computational humanities or cultural anthropology researcher might want to compare the diversity of the Gregorian repertoire to the diversity of, e.g., Irish folk music sessions\footnote{\url{https://thesession.org/}} or modern urban folktales,\footnote{\url{https://digifolk.eu/}} to better understand factors shaping oral transmission. A student of digital humanities might gain appreciation of how historical data may not always reflect textbook definitions of a tradition.
PyCantus --- with sections of documentation specifically for non-chant scholars --- will help such users get started. Furthermore, it provides reference implementations of basic chant processing tasks, thus helping avoid some pitfalls of this domain \citep{lanz2023} and making this huge and fascinating domain available in broader contexts.

\subsection{Article roadmap}

We first review digital chant scholarship to contextualize these contributions (\autoref{sec:relatedwork}). Then we describe the Cantus data ecosystem from which CantusCorpus data is derived (\autoref{sec:cantusecosystem}), and the CantusCorpus~v1.0 dataset itself (\autoref{sec:thedataset}). Next, the PyCantus library is introduced (\autoref{sec:library}), and we show how it allows the integration of further sources of chant data beyond the Cantus ecosystem via a pilot dataset from the digital edition project Corpus Monodicum (\autoref{sec:monodicum}). We conclude with a discussion of the value and limitations of our work (\autoref{sec:conclusions}).

\vspace{3mm}

\section{Related Work}
\label{sec:relatedwork}

How do CantusCorpus v1.0 and PyCantus fit within the broader environment of digital chant scholarship?

\subsection{Sources of digital chant data}

Both CantusCorpus v1.0 and PyCantus are derived from \textbf{the Cantus ``ecosystem''}: the Cantus Database \citep{lacoste2012cantus} and the network of 18 further databases that implement the same data model that is centred around the Cantus Index federated search interface \citep{lacoste2022cantus}. This is by far the largest digital chant data resource; it is continuously growing by tens of thousands of chant records each year. Given its central role, the Cantus ecosystem is discussed in detail in \autoref{sec:cantusecosystem}.

We specifically build on CantusCorpus~v0.2 \citep{cornelissen2020studying}, derived from the Cantus Database. This corpus has been the \textit{de facto} standard for replicable computational chant research, as it has been the only dataset available to date \citep{cornelissen2020mode,cornelissen2020studying,lanz2023,lanz2025chantlab,lanz2025segmentation,hajicjr2025unseen}.
It contains the entirety of the Cantus Database as of 2020, for a total of 497,071 chant records.  
However, it does not collect information about sources, and does not include chants from other databases than Cantus, thus not taking advantage of the breadth of the Cantus ecosystem.

Another major project cataloguing chant digitally is \textbf{Corpus Monodicum}.\footnote{\url{https://www.adwmainz.de/forschung/projekte/corpus-monodicum-die-einstimmige-musik-des-lateinischen-mittelalters-gattungen-werkbestaende-kontexte/informationen.html}} This is a 16-year chant editorial effort that has since 2013 been developing a database\footnote{\url{https://corpus-monodicum.de/}} and digital tools: the Monodi editor \citep{eipert2019editor}, the corresponding MonodiKit library \citep{eipert2023monodikit}, the OMMR4All tool for diastematic optical music recognition \citep{wick2019ommr4all,hartelt2024optical}. It has, so far, accumulated over 5,000 fully transcribed melodies. While this is an order of magnitude smaller than
the Cantus Database, Corpus Monodicum represents a completely independent digital chant scholarship effort, with different objectives and database structure. It thus presents an opportunity to test whether the design of PyCantus and its data model generalise beyond the Cantus ecosystem, and to provide an example for other sites with chant data. 
This pilot ``trial by fire'' for PyCantus is described in \autoref{sec:monodicum}.

There are also limited chant datasets for specific research questions: e.g., the Corpus Troporum Dataset 
\citep{eipert2025corpus,eipert2023tropers}, and the small Christmas and Introits datasets 
\citep{hajic2023,hajicjr2025genomeOfMelody}.

\subsection{Digital resources for contemporary chant singers}

Further chant databases exist that are designed around the needs of contemporary singers and liturgy rather than research.
The most significant is \textbf{GregoBase},\footnote{\url{https://gregobase.selapa.net}} which collects chant repertoire primarily from modern editions of Gregorian chant produced in the Solesmes monastery since the early 20th century for a total of 18,930 chant records.
The database uses the GABC encoding, which can be imported into the Volpiano-based Cantus ecosystem with the \texttt{chant21} library \citep{cornelissen2020studying}. 
The GABC format is also used for typesetting with the Gregorio TeX-based software pipeline.\footnote{\url{https://gregorio-project.github.io/}}
and by the \texttt{scrib.io} editor of the Neumz project.\footnote{\url{https://scrib.io/}} 

The Neumz project includes large-scale recording of contemporary chant singing, totalling over 7,000 hours. This material is being further processed in the Repertorium project.\footnote{\url{https://repertorium.eu/}} 
Digital materials for contemporary liturgy are also available via other apps such as SquareNote,\footnote{https://squarenote.co/} and the \texttt{jgabc} toolkit.\footnote{\url{https://bbloomf.github.io/jgabc/faq.html}}

This combination of digital resources covers liturgical needs well, but is not as relevant for medieval chant scholarship.

\subsection{Computational chant research}
\label{subsec:relatedwork:tools}

There is much recent interest in studying Gregorian chant using computational methods. 
Previous results of manual analysis, necessarily on limited selections of repertoire, can be computationally revisited at much greater scope, and with more transparency in analysis and a correspondingly clearer ``standard of proof'' \citep{McBride2025commentary}.

Most of this research focuses on melody.
The melodic arch hypothesis \citep{narmour1992analysis} has been verified in chant \citep{cornelissen2020studying}.

Previously posited melodic ``dialects'' \citep[p.537-538]{solesmes1960groups, glasenapp2020pray, hornby2002tracts, hiley1993western}, most notably the ``East Frankish'' vs. ``West Frankish'' dialects \citep{wagner1925bericht, blachly1990some}, have been revisited with phylogenetic methods \citep{hajic2023, hajicjr2025genomeOfMelody}.

Centonisation \citep{ferretti1934estetica, levy1970italian}, the idea that Gregorian melodies are constructed from a vocabulary of stable melodic segments, which has been subject to a critical debate \cite[p.~74--75]{treitler1975centonate,hiley1993western}, has recently been empirically shown as highly unlikely using Bayesian non-parametrics \citep{lanz2025segmentation} over entire manuscripts.
However, characteristic melodic patterns do help describe repertoire structure \citep{helsen2021sticky}, or can be used to classify chant melodies by tradition, such as the offertory \citep{kranenburg2017offertories}.
Tools exist for approximate melodic pattern discovery in adiastematic notations (neumes without stafflines) \citep{hornby2022CEAP}, as well as for transitional Aquitanian notation \citep{phan2025analyser}.
Segment-based mode classification using machine learning \citep{cornelissen2020mode,lanz2023,lanz2025segmentation} has been shown to outperform music-theoretical approaches based on the Boethian definition of finals and ranges \citep[ch.1]{atkinson2008critical}, as well as pitch profiles and pitch class profiles \citep{huron2006cognitive}; the link between Gregorian modality and memory of the ``cantus tradition'' \citep[ch.3]{atkinson2008critical} has recently been established empirically \citep{lanz2025segmentation}.

For mapping repertoire, revisiting the idea of grouping sources into traditions \citep{ottosen2008responsories}, network models have been used \citep{eipert2023tropers,dvorakova2025chantmapper} in combination with map visualisation to discover non-trivial repertoire traditions.

Finally, Optical Music Recognition (OMR) for chant has seen significant long-term attention \citep{helsen2014chantOmr,fujinaga2019single,vigliensoni2019image,martinez2023holistic,hartelt2024optical,FuentesMartinez2026amnlt}. However, in relation to our work, OMR exists ``upstream'' of the databases, as a pathway for automated data entry.

\begin{figure*}[t]
    \centering
    \includegraphics[width=1.0\linewidth]{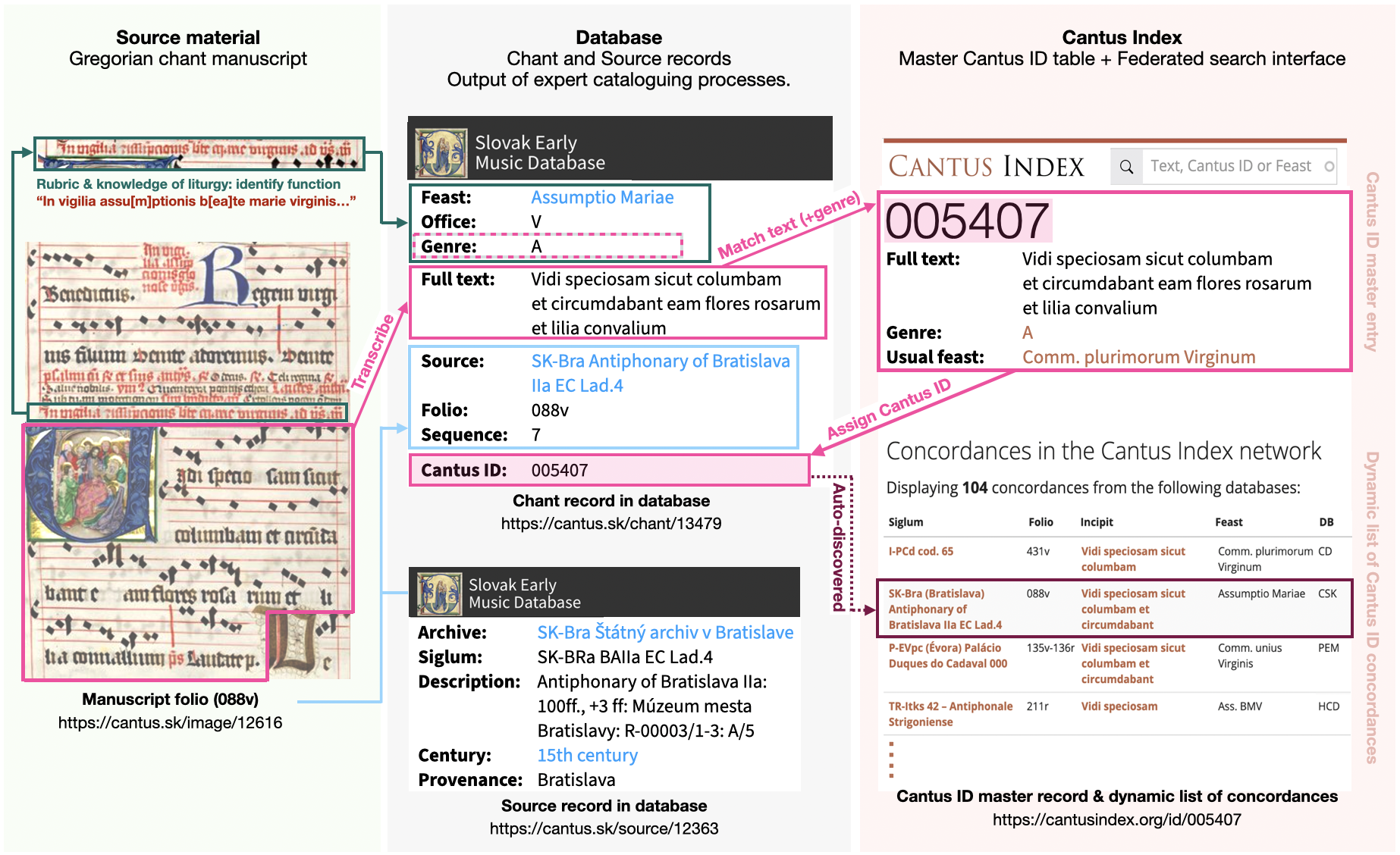}
    \caption{Cataloguing chant in the Cantus ecosystem. An expert identifies chants in a manuscript (left panel), and creates their database records (middle panel, top). The key step in creating a chant record is assigning the Cantus ID: identifying which unit of chant repertoire is on the page (\textcolor{fig3pink}{\textbf{pink}} process). Besides transcribing the text, 
    liturgical expertise is needed, as one must correctly interpret abbreviated notes in the manuscript --- rubrics --- to identify the liturgical position and function of the chant (\textcolor{fig3green}{\textbf{dark green}} process);
    together with the text of the chant, this allows one to select the correct Cantus ID among the `master records' in Cantus Index (right panel, top). A link to the source record (middle panel, bottom) and page (folio) within the source is added (\textcolor{fig3blue}{\textbf{light blue}} process). 
    Once a record with a Cantus ID is added to a database in the Cantus ecosystem, the Cantus Index federated search mechanism (right panel, bottom) will discover the record (\textcolor{fig3purple}{\textbf{dark purple}} process). 
    Descriptions of individual fields mentioned in the figure can be found in \autoref{tab:chants} and \autoref{tab:sources}. (Screenshots from the given URLs have been adjusted for readability.)}
    \label{fig:cataloguing-schema}
\end{figure*}


\section{The Cantus Data Ecosystem and CantusCorpus policy}
\label{sec:cantusecosystem}

The medieval chant resources in the Cantus Index network are indispensable tools for chant research and discovery. The 19 databases and chant research sites connected through the Cantus Index hub have facilitated the traditional study of chant by increasing findability of related materials and by aggregating data in a variety of methods.
In order to understand data in CantusCorpus, a brief overview of the Cantus ecosystem is needed.

The process of cataloguing chant in a Cantus ecosystem database consists of identifying individual units of repertoire in a manuscript, determining their function and position in liturgy, transcribing their text (and optionally melody), and most importantly, assigning the Cantus ID, which allows aggregating catalogue records for the same chant --- unit of the Gregorian repertoire --- across all catalogued sources. This process is shown in \autoref{fig:cataloguing-schema}.

\subsection{History of the Cantus Ecosystem}

The Cantus ecosystem has at its core the metadata schema and principles of the Cantus Database, created forty years ago as ``Cantus: a data base for Gregorian chant.'' 

It started in 1984 with Ruth Steiner and the Cantus Database being developed on a personal computer, later on a mainframe in COBOL, at The Catholic University of America in Washington, D.C..
In 1997, Terence Bailey transferred it to the University of Western Ontario and migrated it to Microsoft Access, enabling richer annotation.

The project grew further. 
In 2011 it was rebuilt in the Drupal system and, crucially, 
the master table of chant texts was split from the manuscript inventory records 
and Cantus Index was built around it as a separate site, intended to be both an online catalogue of all the chant texts and melodies found to date and a federated search of multiple chant resources. 
Cantus Index enabled researchers to build databases with an identity separate from Cantus DB,
while still contributing their data with a shared data model through a single user interface. Thus, Cantus grew from a database into an ecosystem.
Importantly, this growth significantly enlarged the pool of scholars with final authority on prioritising sources for cataloguing, minimising biases such as an implied preference for earlier sources predicated on the 20th-century efforts to reconstruct, in simplified terms, ``the original melodies'' \citep{solesmes1960groups,froger1978critical}.
In 2023, the Cantus Database was migrated to the Python-based Django content management system.

The Cantus ecosystem is an example of success at scale in the Digital Humanities.
At the same time, its complex history means the data is not entirely consistent.
Although the database schema is largely the same across Cantus Index-connected databases, there is considerable flexibility within the requirements defined by the Cantus Index API\footnote{\url{github.com/dact-chant/cantus-index/blob/main/README.md}} in how certain fields are filled, which results in a variety of cataloguing policies in some of the other Cantus Index databases.
Even as the Digital Analysis of Chant Transmission project (DACT\footnote{\url{https://dact-chant.ca/}})
recently formed the Cantus Index Directors’ Council to establish governance frameworks in an otherwise unregulated environment, standardisation among the Cantus network of databases will be a long-term endeavour.

\subsection{Downstream-of-database policy}
\label{sec:harmonisation}

We identified numerous incompatibilities between individual Cantus network databases, both for chants and sources. These mostly stem from a lack of controlled vocabularies, which in turn leaves opportunities for diverging editorial policies.
This openness has been key for sustaining and growing the Cantus network, but it complicates computational work.
Feast spellings are not standardised (\texttt{Agathae, infra oct.} vs. \texttt{Agathae,8}, or \texttt{Adv., hebd. 4., sabbato} vs. \texttt{Sabbato Hebd. 4 Adv.}, which is difficult to discover with e.g.
Levenshtein distance), 
provenances are provided at incompatible levels of granularity (individual church, a city, or a country), as well as centuries of origin (\texttt{13th century}, \texttt{c.1250}, \texttt{late 13th century}, \texttt{possibly around 1225-50}), etc.

Some of these issues are actively being resolved. 
For instance, sigla in the Cantus Database have recently been officially brought into line with RISM policy. The Cantus Database has also published its annotation manual,\footnote{\url{https://cantusdatabase.org/documents/}}
and the Portuguese Early Music Database (PEM) has written guidelines as well.\footnote{\url{https://pemdatabase.eu/node/93860}}
So, one could argue that the right time to publish a dataset such as CantusCorpus is when this standardisation process is done. 
However, this is unrealistic. Given the complexity of the Cantus data ecosystem, 
implementing a controlled vocabulary for feast names, for example, can take months to design and apply across the network, and more phenomena --- such as Old Hispanic rite or chant in vernacular languages --- are being introduced. 

Therefore, rather than waiting for an illusory stable state, we adopt the policy of working \textbf{downstream of the databases}.
We take the position that CantusCorpus cannot sustainably do more than reflect the current consensus on chant editorial policies within the Cantus ecosystem, and should not attempt to supersede the editorial process within Cantus Index governance.
Instead, we take a snapshot, document its limitations, and provide the ETL source code together with the dataset, so that new versions of CantusCorpus will be iteratively available as the databases grow and their standardisation processes proceed. (This is also why the version number must be an integral part of the dataset name.)
We collected the issues in cross-database harmonisation into a document (\hyperref[sup:1]{supplementary file S1}) distributed with CantusCorpus~v1.0. Besides describing v1.0's limitations, it can serve as a roadmap for further standardisation.


\begin{table}[t]
\centering
  \begin{tabular}{lp{5.5cm}}
  \toprule
  \bfseries Field & \bfseries Description \\ \toprule
  \texttt{\textbf{chantlink}}* & URL link directly to the chant entry in the external database. \textbf{Unique ID.} \\ \midrule
  \texttt{incipit}* & The opening words of the chant. \\ \midrule
  \texttt{cantus\_id}* & The Cantus ID associated with the chant (e.g., \texttt{007129a}). \\ \midrule
  \texttt{mode} & Mode of the chant. \\ \midrule
  \texttt{siglum}*  & Abbreviation for the source manuscript or collection (e.g., \texttt{A-ABC Fragm. 1}), ideally RISM.     \\ \midrule
  \texttt{position} & Order of the chant in the office (1st, 2nd, etc.). \\ \midrule
  \texttt{folio}* & Folio information for the chant. \\ \midrule
  \texttt{sequence} & The order of the chant on the folio. \\ \midrule
  \texttt{feast} & Feast or liturgical occasion when the chant is used. \\ \midrule
  \texttt{feast\_code} & Additional identifier unifying feasts with multiple spellings. The values are meaningful in Cantus Index. \\ \midrule
  \texttt{genre} & Genre of the chant, such as antiphon (\texttt{A}), responsory (\texttt{R}), etc. \tablefootnote{See \url{https://cantusindex.org/genre} for genre abbreviations.} \\ \midrule
  \texttt{office} & The liturgy in which the chant is used, such as Matins (\texttt{M}) or Lauds (\texttt{L}). \\ \midrule
  \texttt{srclink}* & URL link to the source in the external database. \\ \midrule
  \texttt{melody\_id} & The Melody ID associated with the chant (e.g., \texttt{001216m1}). Rarely used.\\ \midrule
  \texttt{full\_text} & Full text of the chant. \\ \midrule
  \texttt{melody} & Melody encoded in Volpiano. \\ \midrule
  \texttt{db}* & Abbreviation of the source database. \\ \midrule
  \texttt{image} & URL link to an image of the manuscript page. \\ 
  \bottomrule
  \end{tabular}
  \caption{Chants fields overview. The asterisk (\texttt{*}) indicates required fields.}
\label{tab:chants}
\end{table}

\section{CantusCorpus v1.0}
\label{sec:thedataset}

The CantusCorpus v1.0 dataset collects all records catalogued over the 40 years of the existence of Cantus-style databases that were available through the Cantus Index interface as of May~20th,~2025. 
The dataset has two main components: \textbf{chants} and \textbf{sources}.
A third component is the Extraction-Transformation-Load (ETL) layer: software that was used to collect the dataset.

The \textbf{chants} are the catalogue records of individual items of the Gregorian repertoire, such as antiphons, responsories, alleluias, etc., with their catalogue records according to the Cantus Index standard -- especially the Cantus ID (see \autoref{fig:cataloguing-schema}). These records also have a unique identification in chantlink -- the URL of the record in their home database. 

The \textbf{sources} represent liturgical manuscripts, early printed sources, and manuscript fragments that contain the chants.
Source records contain the fundamental music-historical metadata in CantusCorpus:
provenance (location where the book was likely written and/or used),
the century when it was likely written, 
and its \textit{cursus} --- what type of ecclesiastical institution it belonged to (though not all databases support this field).
While chant records are relatively unambiguous, 
source metadata can be uncertain, depending on the musicological situation of individual sources.

\subsection{Chants}
\label{subsec:chants}

The overview of the fields in the chants CSV file is given in \autoref{tab:chants}. It follows the Cantus Index JSON API.\footnote{\url{https://github.com/dact-chant/cantus-index/blob/main/README.md}}

\subsubsection{Obtaining the chants}
\label{sec:chants:obtain}
All records of the chants in CantusCorpus~v1.0 were obtained by scraping the JSON endpoint of the Cantus Index, which is commonly used to obtain concordances.
For each genre, we gathered its Cantus IDs and collected them sequentially via the endpoint URL.\footnote{\url{https://cantusindex.org/json-cid/{CantusID}}} 
The code can be found in the dataset repository.\footnote{\url{https://github.com/dact-chant/CantusCorpus/tree/main/cantuscorpus_1.0/scraping}}

While 19 databases are listed as part of the Cantus Index ecosystem, after traversing all Cantus IDs via the JSON API, we found chant records only from the following 10 databases:

\begin{itemize}[noitemsep,leftmargin=*]
    \item Cantus Database\footnote{\url{https://cantusdatabase.org/}} (CD)
    \item Medieval Music Manuscripts Online\footnote{\url{https://musmed.eu/}} (MMMO)
    \item Slovak Early Music Database \footnote{\url{https://cantus.sk/}} (CSK)
    \item Cantus Fontes Bohemiae\footnote{\url{https://cantusbohemiae.cz/}} (FCB)
    \item Cantus Planus in Polonia\footnote{\url{https://cantusplanus.pl/}} (CPL)
    \item Portuguese Early Music Database\footnote{\url{https://pemdatabase.eu/}} (PEM)
    \item Spanish Early Music Manuscripts Database\footnote{\url{https://musicahispanica.eu/}} (SEMM)
    \item Hungarian Chant Database\footnote{\url{https://hun-chant.eu/}} (HCD)
    \item Medieval Music Manuscripts from Austrian Monasteries\footnote{\url{https://austriamanus.org/}} (A4M)
    \item Codicologica et Hymnologica Bohemica Liturgica\footnote{\url{https://hymnologica.cz/}} (HYM)
\end{itemize}

\subsubsection{Data cleaning}
\label{subsec:chants:clean}

After converting the scraped JSON files to more compact CSVs (``Excel flavour'') for each genre and joining them into a single file, we performed the following cleanup steps:

\begin{itemize}[noitemsep,leftmargin=*]
    \item Standardise \texttt{genre} based on the genre list through which the chant record was found.\footnote{\url{https://cantusindex.org/genre}}
    
    \item Discard duplicate records of the same chants collected from different genre lists (using the \texttt{chantlink} as a unique identifier of the chant record).
    
    \item Exclude chants from sources that were themselves not publicly visible (see \autoref{subsec:sources:describ}).
    
    \item Rewrite \texttt{office} values for chant records from the Hungarian Chant Database based on its web interface. 
\end{itemize}

\noindent More details about these steps of chant data cleaning can be found in \hyperref[sup:4]{supplementary file S4}.

\subsection{Sources}
\label{subsec:sources}

The situation for sources is much less straightforward than for chants, because Cantus Index performs no aggregation of sources and thus has no JSON endpoint, unlike for chants.

\subsubsection{Obtaining the sources} 
\label{subsec:sources:obtain}

The information available in each chant record is limited to the URL of the source on its home database webpage and its siglum.
The webpage exposes some subset of source metadata. 
We implemented scrapers for front-ends of all ten databases whose chants were available via the Cantus Index JSON endpoint.

\subsubsection{Sources data fields}
\label{subsec:sources:describ}

\begin{figure*}
    \centering
    \includegraphics[width=0.99\linewidth]{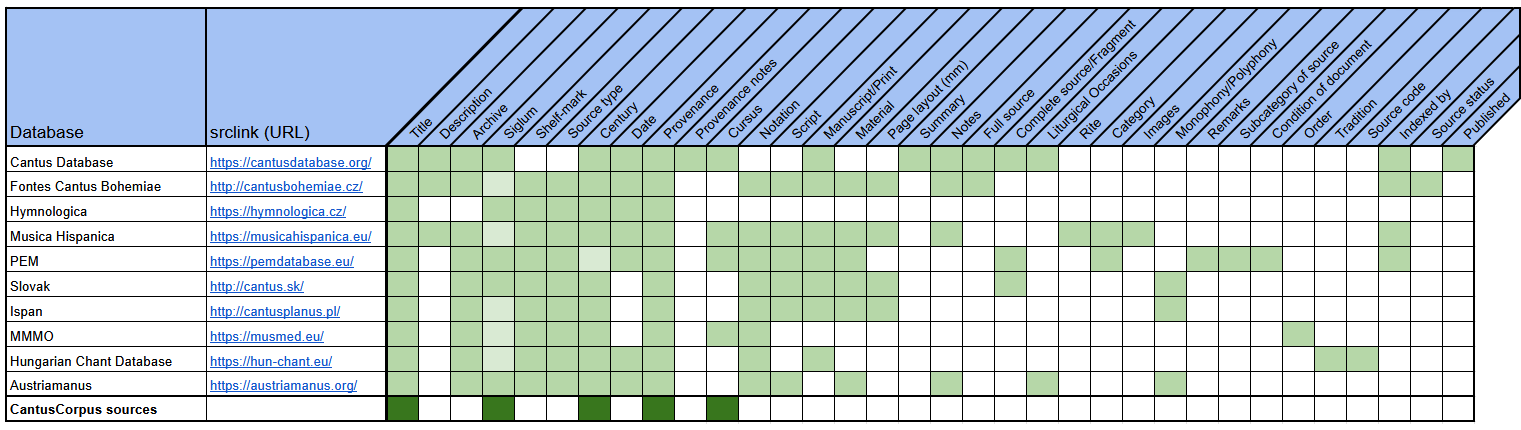}
    \caption{Overview of support for source metadata among CI database front-ends. Lightest green indicates support under a differently named field, darkest indicates fields that were selected to be included in CantusCorpus~v1.0.}
    \label{fig:source-fields}
\end{figure*}

It is mainly due to the ``self-documenting'' nature of the Gregorian chant that a fairly unified data schema emerged for the chant data. 
However, information about sources is not collected consistently among the Cantus Index databases. 
We surveyed what fields they provide 
highlighting the variability of available data in \autoref{fig:source-fields}.

We selected those fields that are
to be found in all databases (see \autoref{tab:sources}) and then added \texttt{cursus}. We consider \textit{cursus} important since it relates to repertoire structure, 
and it is often known (and can then usually be found in the textual source descriptions, even if not annotated as a field).
Then we automatically derived value \texttt{num\_century} from the obtained \texttt{century} value.

\begin{table}[H]
\centering
  \begin{tabular}{lp{5.5cm}}
  \toprule
  \bfseries Field & \bfseries Description \\ \toprule
  \texttt{title} & Manuscript name (may use siglum) \\ \midrule
  \texttt{siglum}*  & Abbreviation for the source manuscript, possibly RISM.     \\ \midrule
  \texttt{century} & Text identifying the century of the source. \\ \midrule
  \texttt{provenance} & Place of origin or use of the source. \\ \midrule
  \textbf{\texttt{srclink}*} & URL link to the source in the external database. \textbf{Unique ID.}\\ \midrule
  \texttt{cursus} & Secular or Monastic \textit{cursus} of the source. \\\midrule
  \texttt{num\_century} & Integer representation of century. \\
  \bottomrule
  \end{tabular}
  \caption{Sources fields overview. The asterisk (*) indicates required fields.}
\label{tab:sources}
\end{table}

\subsubsection{Data cleaning}
\label{subsec:sources:clean}

The source records were compiled into a CSV file (``Excel flavour'') and a series of data refinement procedures was implemented.
Because there are orders of magnitude fewer sources than chants, this could involve manual steps.

\begin{itemize}[noitemsep,leftmargin=*]
    \item Resolve changes in \texttt{http} vs \texttt{https} that did not propagate through scraping because of working redirects.
    
    \item Identify and inspect sources 
    for which metadata scraping was unsuccessful -- some added manually (missing but restorable \texttt{siglum}), some discarded, together with chants that reference them (\texttt{srclink} URL returns \textit{Access denied}).
    
    \item Identify duplicated sources based on siglum, check them one by one 
    and resolve each situation, e.g. by discarding one of them (and its chant records) or unifying parts of one manuscript under one \texttt{srclink} while keeping the intersection of chant records only once.
    
    \item Automatically derive the \texttt{num\_century} value from the \texttt{century} string.
\end{itemize}

\noindent
More details about the steps that source data cleaning underwent can be found in the \hyperref[sup:4]{supplementary file S4}.

\subsection{CantusCorpus v1.0 statistics} 
\label{subsec:sources:stats}

Tables \ref{tab:chants:stats} and \ref{tab:sources:stats} summarise the counts of chant records and melodies, and source records and their completeness. Compared to cataloguing chant texts (many of which are Biblical or otherwise known and standard in the repertoire), transcribing melodies is slow and difficult, and thus remains a minority practice. 
A small number of sources being annotated with \textit{cursus} values reflects the number of databases that implement this field (see \autoref{fig:source-fields}).

Table \ref{tab:db_stats} brings a broader overview of the distribution of data among the source databases. 
The Cantus Database, by far the oldest part of the network, is (still) the largest (just under half of all records).
Musica Hispanica (SEMM) looks 
``most distinct'' among the others, having the highest number of uniquely appearing Cantus IDs.
The surprisingly high ratio of unique Cantus IDs in the Codicologica et Hymnologica database (HYM) comes from language-specific Cantus IDs for chants written in vernacular (Czech). 
Note that CSK and A4M have many fragments or incomplete manuscripts. 

Further statistics can be found in the \hyperref[sup:3]{supplementary file S3}.

\begin{table}[H]
\centering
  \begin{tabular}{p{5.5cm} >{\raggedright\arraybackslash}p{1.4cm}}
  \toprule
  \bfseries  Chant records in chants.csv & \bfseries Number \\ \midrule
  All & \parbox[t]{\linewidth}{\raggedleft 888010} \\
  With Volpiano melody  & \parbox[t]{\linewidth}{\raggedleft 60588}   \\
  With Volpiano melody of 20+ notes  & \parbox[t]{\linewidth}{\raggedleft 44625}  \\
  \bottomrule
  \end{tabular}
  \caption{Basic quantitative values of the chants part of the CantusCorpus v1.0 dataset.}
\label{tab:chants:stats}
\end{table}

\begin{table}[H]
\centering
  \begin{tabular}{p{5.5cm} >{\raggedright\arraybackslash}p{1.4cm}}
  \toprule
  \bfseries  Source records in sources.csv & \bfseries Number \\ \midrule
  All & \parbox[t]{\linewidth}{\raggedleft 2278}      \\
  All with 100+ chants & \parbox[t]{\linewidth}{\raggedleft 508}      \\
  Those with provenance value & \parbox[t]{\linewidth}{\raggedleft 1606}    \\
  Those with century value & \parbox[t]{\linewidth}{\raggedleft  2240}    \\
  Those with \textit{cursus} value & \parbox[t]{\linewidth}{\raggedleft  345}    \\
  \bottomrule
  \end{tabular}
  \caption{Basic quantitative values of the sources part of the CantusCorpus v1.0 dataset.}
\label{tab:sources:stats}
\end{table}

\begin{table*}
\centering
\begin{tabular}{p{2.9cm} >{\raggedleft\arraybackslash}p{1.9cm} >{\raggedleft\arraybackslash}p{1.5cm} >{\raggedleft\arraybackslash}p{2.7cm} >{\raggedleft\arraybackslash}p{1.9cm} >{\raggedleft\arraybackslash}p{3.5cm}}
\toprule
\bfseries Source DB code & \bfseries \# chants & \bfseries \# CIDs & \bfseries \# unique CIDs & \bfseries \# sources & \bfseries \# sources (100+) \\ \midrule
 CD & 429982 & 30350 & 14662 & 231 & 166 \\ \midrule
 MMMO & 212231 & 17479 & 7503 & 426 & 151 \\ \midrule
 CSK & 22539 & 7201 & 212 & 542 & 12 \\ \midrule
 FCB & 36103 &7889 & 534 & 30 & 29 \\ \midrule
 CPL & 30433 & 7666 & 143 & 27 & 17 \\ \midrule
 PEM & 32738 & 9184 & 538 & 305 & 25 \\ \midrule
 SEMM & 104678 & 23103 & 11625 & 487 & 81 \\ \midrule
 HCD & 11278 & 5374 & 54 & 10 & 9 \\ \midrule
 A4M & 2738 & 2006 & 12 & 142 & 3 \\ \midrule
 HYM & 5290 & 680 & 323 & 83 & 20 \\
\bottomrule
\end{tabular}
\caption{Overview of data distribution among source databases. The symbol \texttt{\#} is used as an abbreviation for ``number of''. Abbreviations of database codes can be found in \autoref{sec:chants:obtain}. The column annotated with \texttt{\# sources (100+)} contains the number of sources with more than 100 chant records associated with them.}
\label{tab:db_stats}
\end{table*}


\section{The PyCantus Library}
\label{sec:library}

PyCantus is a lightweight Python library for loading and manipulating the CantusCorpus v1.0 dataset (see \autoref{sec:thedataset}). 
The ``division of labour'' between the library and dataset is shown in \autoref{fig:simple_schema}.
Most importantly, PyCantus implements a data model for CantusCorpus v1.0. 
However, the library is an independent component, so the data model can be reused for datasets assembled from other sources of chant data. 
PyCantus functionality is introduced by tutorials (see \autoref{subsec:library:tutorials}). 
The library is available at \url{https://github.com/dact-chant/PyCantus/}.

\subsection{Data Model}

\begin{figure}[]
    \centering
    \includegraphics[width=0.87\linewidth]{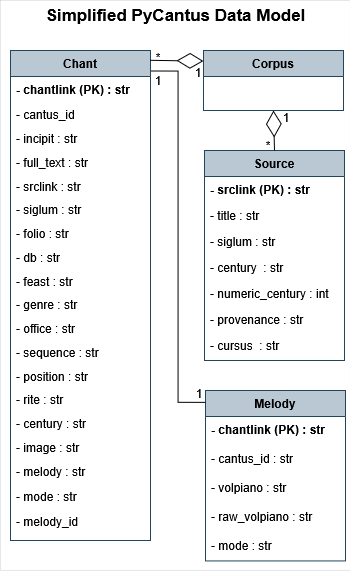}
    \caption{Simplified schema of the PyCantus data model (``content'' attributes only). Full UML model can be found in \hyperref[sup:2]{supplementary file S2}.}
    \label{fig:data_model}
\end{figure}

The core of PyCantus is the data model (see \autoref{fig:data_model}). The \texttt{Chant} and \texttt{Source} classes represent the corresponding items of the dataset,
the \texttt{Corpus} class aggregates them, and a \texttt{Melody} class supports abstracting away from specific melody encodings in the future.

Each data class has further convenience methods, such as \texttt{Corpus} exporting itself into a CSV file, or \texttt{Melody} cleaning Volpiano representations.

To prevent accidental data manipulations, a lock can be set 
in the \texttt{Corpus} that limits access to attribute setters of \texttt{Chant}, \texttt{Source} and \texttt{Melody} objects in \texttt{Corpus}.

\subsection{Loaders}

Data are loaded into the \texttt{Corpus} class via a \texttt{CsvLoader} object.
The loader also handles 
validation. Given the lack of controlled vocabularies (see \autoref{sec:harmonisation}), it currently cannot do more than verify whether mandatory fields and their values are present. 

\subsection{Replicable Filtering}
\label{subsec:filter}
Most experimental workflows involve filtering of the input chants: by genre, feast set, provenance, etc. Accurately describing these filters is essential for research reproducibility. 
Providing the filtered datasets themselves is an incomplete solution: follow-up work often should assess how findings apply to different sets of repertoire selected according to the same principles. 

To this end, PyCantus provides the \texttt{Filter} class, which can export filtering conditions to a YAML configuration file (static and easy-to-share), and apply such a file to a corpus. 
We also provide a web app to build these files.\footnote{Deployed: \url{https://filterforpycantus.owx.cz/filter/}}\footnote{Django template source code: \url{https://github.com/DvorakovaA/filterforpycantus}}

\subsection{Corpus operations history}
To further promote reproducibility, PyCantus tracks the history of data operations on each \texttt{Corpus} class. This record can be exported as human-readable text, guiding the user through applied operations so they can easily replicate them in their own code.

\subsection{Documentation for novices to chant}
Two types of users might be interested in PyCantus:
chant experts exploring computational methods, but also
digital humanities scholars or students exploring Gregorian chant.
We therefore provide not only technical and user documentation, but also a document for people new to Gregorian chant. It provides a basic introduction to the material, 
including computing approaches, 
and so reduces the entry barrier.

\subsection{Tutorials}
\label{subsec:library:tutorials}
Finally, we provide tutorials: two introductory (basic PyCantus concepts, and basics of working with chant data), and three that reproduce recent chant computing experiments on the CantusCorpus~v1.0 dataset: manuscript ``community'' detection \citep{eipert2023tropers, dvorakova2025chantmapper},  estimating the number of unseen chants \citep{hajicjr2025unseen, moss2025rest}, and classifying melodies into modes by segments \citep{cornelissen2020mode}. The value of CantusCorpus~v1.0 has been clearly demonstrated: reproducing the unseen chants experiment on CantusCorpus~v1.0 rather than the more limited CantusCorpus~v0.2 
shows that some of the discussion in \citep{hajicjr2025unseen} was an artefact of limited Mass data in CantusCorpus~v0.2.


\section{Beyond Cantus: integration of Corpus Monodicum data}
\label{sec:monodicum}

PyCantus was designed to make other sources of chant data interoperable with those derived from the Cantus ecosystem. 
To demonstrate this capability, and to provide actionable insight into how such a process would work, 
we built a pilot PyCantus-compatible dataset from the data of Corpus Monodicum. 

\subsection{Corpus Monodicum}

Corpus Monodicum\footnote{\url{https://corpus-monodicum.de/}} (CM) is a project that is creating editions of sacred and secular monophonic chant repertories. 
It encodes chants in MEI.
As a long-term project of 16 years, it has produced significant computing infrastructure of its own: the MonodiKit software library \citep{eipert2023monodikit}, the Monodi+ editor \citep{eipert2019editor}, and the OMMR4All recognition software \citep{wick2019ommr4all,hartelt2024optical}. 
It has an editorial history, policy, and codebase completely independent of the Cantus ecosystem. Also, there is minimal overlap in the catalogued chants, and the focus on Germany could add a lot of valuable material, analogous to the value of SEMM for Iberian material, to the overall corpus.
Thus, it serves as a good ``test case'' for whether the PyCantus data model can serve as a shared infrastructure for chant computing in general: if CM data can be made available as a PyCantus dataset, the chance that \textit{any} chant data plausibly can goes up.

The primary challenge
is that some metadata fields do not align between the two projects. One field is critical: the Cantus ID. This identifier is mandatory in PyCantus but is not recorded in Corpus Monodicum.

\subsection{Pilot ETL process}

The CM dataset is available from the Open Science Foundation Repository.\footnote{\url{https://osf.io/mfpkd/}}
The pilot sub-corpus comprises all Proper Mass antiphons --- CM nomenclature for chants labelled \texttt{In}, \texttt{Of}, and \texttt{Co} in Cantus Index --- of CM, excluding the Graduale Synopticum (GS) sub-corpus that is built from an edition rather than medieval sources. 
We chose to start with this subcorpus due to its relative standardisation and familiarity. 
Our aim is to generate a CSV file that PyCantus can import without issues.

\noindent Our ETL process is as follows:

\begin{enumerate}
    \item \textbf{Load the CM dataset into \texttt{MonodiKit} and select the sub-corpus.} 
    \texttt{MonodiKit}’s default workflow returns a list of document objects that can be accessed directly from Python.
    \item \textbf{Find Cantus IDs.} 
    For each document we ran a fuzzy string match (\textit{token\_sort\_ratio} method from the Python library \texttt{rapidfuzz}, with a threshold of 60, which was selected based on preliminary testing) against the list of Proper antiphons of Cantus Index that matched its genre.
    Frequent divergences were caused by different normalisation approaches, such as \textit{hierusalem} and \textit{jerusalem}, or by the inclusion of words such as \textit{alleluia} before \textit{dilexisti iustitiam et odisti...}.
    To avoid false positives, the Cantus ID of the match was stored only when the similarity exceeded the selected threshold.
    When multiple results were returned, a similar fuzzy search was performed for feast days. 
    We then manually 
    confirmed the matched records were correct.
    These records and matching decisions were written to an auxiliary file for manual verification. 
    Note that since we cannot know the exact intersection of CM and Cantus Proper chants, we cannot evaluate how many false negatives 
    this approach might yield.
    
    \item \textbf{Normalise.} Using the inbuilt \texttt{MonodiKit} method, we converted melodies to Volpiano. We then replaced CM sigla with their RISM equivalents using manually compiled concordances.
    
    \item \textbf{Export to PyCantus schema.} We collected the required fields, aligned them with the PyCantus schema, and wrote the result to a CSV file.
\end{enumerate}

\noindent
For sources, the metadata required by PyCantus is directly available in CM, so the sources CSV file can be directly exported.
All ETL code is provided with the dataset.\footnote{\url{https://github.com/corpus-monodicum/CM-to-PyCantus-Pipeline/tree/master}}

The CM dataset contains 498 Proper chants (out 5474 total chants), of which 420 chants from 32 sources were matched with Cantus IDs and thus made it into the pilot PyCantus-compatible dataset.
The remaining 78 chants could not be found, possibly due to limitations of the string matching algorithm, or because the chants may require adding new Cantus IDs.

\subsection{Outcomes}

This workflow exemplifies how PyCantus can be a hub for connecting existing and future chant datasets, like Corpus Troporum, GregoBase, Graduale Synopticum, or the Comparatio Project.\footnote{\url{https://comparatio.irht.cnrs.fr/}} There are a variety of reasons why these individual projects cannot have the same data standards --- but for computational research, they can align with the PyCantus standard to become parts of a larger chant data ecosystem beyond Cantus. 



\section{Conclusions and Future Work} 
\label{sec:conclusions}

Built ``on the shoulders of giants'', that is on the Cantus ecosystem of Gregorian chant data, the CantusCorpus~v1.0 dataset and PyCantus library are the next steps towards large-scale digital (and computational) chant research. 
This infrastructure decouples the four decades' worth of data from the individual databases,
making it easy to work with this valuable data programmatically. 
Thus we lower barriers to entry for digital humanities researchers, reduce friction in collaborations with statisticians and other technical fields, and enhance reproducibility in experimental chant research. 
We especially believe that making chant data easy to use both in research and in educational settings,
and thus making it a more attractive topic from a digital humanities perspective, is 
also an important factor for the sustainability of chant scholarship itself. 

Both the dataset and the library have limitations. The sorest point is the lack of data validation. This follows from the lack of controlled vocabularies for many important fields (feast, office, provenance, etc. --- see \autoref{sec:harmonisation}) in the databases themselves. The validation component will have to be developed together with future major versions of CantusCorpus. 

Requiring Cantus IDs places a burden on non-Cantus databases  converting their data to PyCantus--compatible datasets. A Cantus ID fuzzy matcher would help. A proof-of-concept of this has been developed for the Corpus Monodicum pilot data integration (\autoref{sec:monodicum}); we intend to develop it further and integrate it into PyCantus.

In the PyCantus library, the mechanism of unique identifiers for chants currently follows the Cantus Index convention of providing a unique chant link. This should be made more general as more chant data sources are integrated. To avoid accidental duplicate entries for the same manuscript from different databases, a similar ``soft'' record matcher to highlight suspicious record pairs should be developed as well.
Support for non-Volpiano melody encodings (primarily GABC) should also be implemented.

Digital musicology is a perspective that may significantly help answer large-scale questions in chant scholarship about the structures of diversity and processes of change within the widespread medieval Gregorian chant tradition. 
Appropriate research infrastructure is essential for this approach. We believe that the PyCantus library and CantusCorpus v1.0 will become important elements of that infrastructure, enabling the vast collection of digitally catalogued chants collected over the last four decades to yield knowledge in new and exciting ways.


\section*{Additional Files}
\textbf{Supplementary file S1}\label{sup:1}:

\noindent \textit{Report\_on\_harmonization\_issues.pdf}
-- Overview of encountered issues in the dataset composition process.
\url{https://github.com/dact-chant/CantusCorpus/blob/main/cantuscorpus_1.0/Report_on_harmonization_issues.pdf}

\vspace{1mm}

\noindent \textbf{Supplementary file S2}\label{sup:2}: 

\noindent \textit{pycantus\_data\_model\_full\_UML.png}
-- Full PyCantus data model schema in UML. \url{https://github.com/dact-chant/PyCantus/blob/main/docs/source/_static/img/data_model_pycantus_full_uml.png}

\noindent \textbf{Supplementary file S3}\label{sup:3}: 

\noindent \textit{cantuscorpus\_statistics.pdf}
-- Additional plots and count overviews regarding CantusCorpus v1.0.
\url{https://github.com/dact-chant/CantusCorpus/blob/main/cantuscorpus_1.0/dataset_stats.pdf}

\noindent \textbf{Supplementary file S4}\label{sup:4}: 

\noindent \textit{get\_dataset\_from\_scrapes.pdf}
-- File documenting all steps undertaken to prepare CantusCorpus v1.0 dataset from collected data.
\url{https://github.com/dact-chant/CantusCorpus/blob/main/cantuscorpus_1.0/get_dataset_from_scrapes.pdf}

\section*{Data Accessibility Statement}

\noindent The CantusCorpus~v1.0 dataset is available from: \newline
\url{http://hdl.handle.net/11234/1-6041}.

\vspace{2mm} 

\noindent The pilot Corpus Monodicum dataset with accompanying code is available here: \url{https://github.com/corpus-monodicum/CM-to-PyCantus-Pipeline/tree/master}.

\section*{Source code}
\label{sec:code}
The PyCantus library code is available here: \url{https://github.com/dact-chant/PyCantus} .

\vspace{3mm}
The auxiliary Django template and minimal web app for PyCantus creating filter files non-programmatically is available here: \url{https://github.com/DvorakovaA/filterforpycantus/tree/main}

\noindent 

\section*{Competing interests}

The authors declare that they have no competing interests.

\IfFileExists{\jobname.ent}{
   \theendnotes
}{
}

\section*{Acknowledgements}

This work was supported by the Digital Analysis of Chant Transmission (DACT) project, funded by a Partnership Grant from the Social Sciences and Humanities Research Council of Canada (895-2023-1002).
JH further acknowledges support of the project ``Human-centred AI for a Sustainable and Adaptive Society'' (reg. no.: CZ.02.01.01/00/23\_025/0008691), co-funded by the European Union. 
Parts of the datasets and computational approaches were made possible through the ongoing work of Corpus Monodicum, a project of the Academy of Sciences and Literature, Mainz (\url{https://www.adwmainz.de/}), funded by the German Federal and Bavarian State governments.
The work described herein has also been using computational resources provided by the LINDAT/CLARIAH-CZ Research Infrastructure (\url{https://lindat.cz}), supported by the Ministry of Education, Youth and Sports of the Czech Republic (Project No. LM2023062). 

We further extend our thanks to Jennifer Bain, Anna de Bakker, Ichiro Fujinaga and Elsa de Luca for helpful discussions.

\section*{Author contributions}

\textbf{AD:} Conceptualisation, Data curation, Formal Analysis, Investigation, Methodology, Software, Validation, Visualization, Writing -- Original draft, Writing -- review \&~editing

\noindent \textbf{TE:} Data curation, Formal Analysis, Investigation, Resources, Software, Writing -- Original draft, Writing -- review \&~editing

\noindent \textbf{DL:} Investigation, Funding acquisition, Resources, Supervision, Writing -- Original draft, Writing -- review \&~editing

\noindent \textbf{JH:} Conceptualisation, Funding acquisition, Methodology, Project administration, Resources, Software, Supervision,  Validation, Writing -- Original draft, Writing -- review \&~editing


\bibliographystyle{unsrtnat}
\bibliography{bibliography}

@inproceedings{hajic2023,
  author       = {Hajič jr., Jan and Ballen, Gustavo A. and Mühlová, Klára Hedvika and Vlhová-Wörner, Hana},
  title        = {{Towards Building a Phylogeny of Gregorian Chant Melodies}},
  booktitle    = {{Proceedings of the 24th International Society for Music Information Retrieval Conference}},
  year         = 2023,
  pages        = {571-578},
  publisher    = {ISMIR},
  month        = dec,
  venue        = {Milan, Italy},
  doi          = {10.5281/zenodo.10340442},
  url          = {https://doi.org/10.5281/zenodo.10340442}
}

@inproceedings{lanz2023,
  title={Text boundaries do not provide a better segmentation of {G}regorian antiphons},
  author={Lanz, Vojt{\v{e}}ch and Haji{\v{c}} jr., Jan},
  booktitle={Proceedings of the 10th International Conference on Digital Libraries for Musicology},
  pages={72--76},
  year={2023}
}

@article{helsen2021sticky,
  title={The {Sticky} {Riff}: {Quantifying} the {Melodic} {Identities} of {Medieval} {Modes}},
  author={Helsen, Kate and Daley, Mark and Schindler, Jake},
  journal={Empirical Musicology Review},
  volume={16},
  number={2},
  pages={312--325},
  year={2021}
}

@book{ottosen2008responsories,
  title={The responsories and versicles of the Latin office of the dead},
  author={Ottosen, Knud},
  year={2007},
  publisher={BoD--Books on Demand, GmbH},
  isbn={978-87-7691-186-7}
}

@book{glasenapp2020pray,
  title={To Pray without Ceasing: A Diachronic History of Cistercian Chant in the Beaupr{\'e} Antiphoner (Baltimore, Walters Art Museum, W. 759--762)},
  author={Glasenapp, John},
  year={2020},
  publisher={Columbia University}
}

@inproceedings{cornelissen2020mode,
  title={Mode classification and natural units in plainchant.},
  author={Cornelissen, Bas and Zuidema, Willem H and Burgoyne, John Ashley and others},
  booktitle={Proceedings of the 21st International Society for Music Information Retrieval Conference},
  pages={869--875},
  year={2020},
  address={Montreal, Canada}
}

@incollection{lacoste2022cantus,
    author = {Lacoste, Debra},
    isbn = {9780190945442},
    title = "{The Cantus Database and Cantus Index Network}",
    booktitle = "{The Oxford Handbook of Music and Corpus Studies}",
    publisher = {Oxford University Press},
    doi = {10.1093/oxfordhb/9780190945442.013.18},
    url = {https://doi.org/10.1093/oxfordhb/9780190945442.013.18},
    year = {2022}
}

@inproceedings{cornelissen2020studying,
  title={Studying large plainchant corpora using chant21},
  author={Cornelissen, Bas and Zuidema, Willem and Burgoyne, John Ashley},
  booktitle={7th International Conference on Digital Libraries for Musicology},
  pages={40--44},
  year={2020}
}

@article{treitler1975centonate,
  title={"{C}entonate" Chant: "{Ü}bles {F}lickwerk" or "{E} pluribus unus"?},
  author={Treitler, Leo},
  journal={Journal of the American Musicological Society},
  volume={28},
  number={1},
  pages={1--23},
  year={1975},
  publisher={University of California Press on behalf of the American Musicological Society}
}

@book{hiley1993western,
  title={Western {Plainchant}: {A} {Handbook}},
  author={Hiley, David},
  year={1993},
  publisher={Clarendon Press},
  address={Oxford, United Kingdom}
}

@article{blachly1990some,
  title={Some {O}bservations on the "{G}ermanic" {P}lainchant {T}radition},
  journal = {Current Musicology},
  number = {45-47},
  author={Blachly, Alexander},
  publisher = {New York: Department of Music, Columbia University},
  year={1990},
  pages={85-117}
}

@inproceedings{wagner1925bericht,
    author = {Wagner, Peter},
    title = {{G}ermanisches und {R}omanisches im frühmittelalterlichen {K}irchengesang},
    booktitle = {Bericht über den I. Musikwissenschaftlichen Kongreß der deutschen Musikgesellschaft in Leipzig 1925},
    pages = {21-34},
    year = {1925},
}

@book{ferretti1934estetica,
  title={Estetica gregoriana: ossia, {T}rattato delle forme musicali del canto gregoriano},
  author={Ferretti, P.M.},
  number={sv. 1},
  lccn={34015652},
  series={Estetica gregoriana: ossia, Trattato delle forme musicali del canto gregoriano},
  url={https://books.google.cz/books?id=vOWCnQEACAAJ},
  year={1934},
  publisher={Pontificio istituto di musica sacra}
}

@article{froger1978critical,
  title={The critical edition of the {R}oman {G}radual by the monks of {S}olesmes},
  author={Froger, Dom Jacques},
  journal={Journal of the Plainsong \& Mediaeval Music Society},
  volume={1},
  pages={81--97},
  year={1978},
  publisher={Cambridge University Press}
}

@inproceedings{kranenburg2017offertories,
  author       = {Peter van Kranenburg and
                  Geert Maessen},
  title        = {{Comparing Offertory Melodies of Five Medieval Christian Chant Traditions}},
  booktitle    = {Proceedings of the 18th International Society for Music Information Retrieval Conference},
  pages        = {204--210},
  year         = {2017},
  url          = {https://ismir2017.smcnus.org/wp-content/uploads/2017/10/174\_Paper.pdf},
}

@article{lacoste2012cantus,
  title={The {C}antus {D}atabase: Mining for {M}edieval {C}hant {T}raditions},
  author={Lacoste, Debra},
  journal={Digital Medievalist},
  volume={7},
  year={2012},
  publisher={The Open Library Of Humanities}
}

@inproceedings{eipert2023tropers,
    author = {Eipert, Tim and Moss, Fabian C.},
    title = {Poster: Communities in {Medieval} {Troper} {Networks} are {Shaped} by {Carolingian} {Politics}},
    booktitle = {Proceedings of the 10th International Conference on Digital Libraries for Musicology},
    year = {2023}
}

@inproceedings{eipert2023monodikit,
  title={MonodiKit: A data model and toolkit for medieval monophonic chant},
  author={Eipert, Tim and Moss, Fabian C},
  booktitle={Proceedings of the 10th International Conference on Digital Libraries for Musicology},
  pages={67--71},
  year={2023}
}

@inproceedings{vigliensoni2019image,
  title={From image to encoding: Full optical music recognition of {Medieval and Renaissance} music},
  author={Vigliensoni, Gabriel and Daigle, Alex and Liu, Eric and Calvo-Zaragoza, Jorge and Regimbal, Juliette and Nguyen, Minh Anh and Baxter, Noah and McLennan, Zo{\'e} and Fujinaga, Ichiro},
  booktitle={Music Encoding Conference 2019 Book of Abstracts},
  year={2019},
  url = {https://music-encoding.org/conference/2019/abstracts_mec2019/vigliensoni19from%20camera%20ready.pdf}
}

@inproceedings{fujinaga2019single,
  title={Single {Interface} for {Music} {Score} {Searching} and {Analysis} ({SIMSSA}) {Project}: {Optical} {Music} {Recognition} {Workflow} for {Neume} {Notation}},
  author={Fujinaga, Ichiro},
  booktitle={Proceedings of the Computers and the Humanities Symposium (Jin-MonCom). Osaka, Japan: Information Processing Society of Japan},
  pages={281--286},
  year={2019}
}

@article{hartelt2024optical,
  title={Optical {Medieval} {Music} {Recognition} — {A} {Complete} {Pipeline} for {Historic} {Chants}.},
  author={Hartelt, Alexander and Eipert, Tim and Puppe, Frank},
  journal={Applied Sciences (2076-3417)},
  volume={14},
  number={16},
  year={2024}
}

@book{atkinson2008critical,
  title={The critical nexus: Tone-system, mode, and notation in early medieval music},
  author={Atkinson, Charles M},
  year={2008},
  publisher={Oxford University Press}
}

@article{helsen2014chantOmr,
    author = {Helsen, Kate and Bain, Jennifer and Fujinaga, Ichiro and Hankinson, Andrew and Lacoste, Debra},
    title = {Optical music recognition and manuscript chant sources},
    journal = {Early Music},
    volume = {42},
    number = {4},
    pages = {555-558},
    year = {2014},
    month = {10},
    issn = {0306-1078},
    doi = {10.1093/em/cau092},
    url = {https://doi.org/10.1093/em/cau092},
    eprint = {https://academic.oup.com/em/article-pdf/42/4/555/9676607/cau092.pdf},
}

@book{hornby2002tracts,
  title={Gregorian and {Old} {Roman} {Eighth-Mode} {Tracts}: {A} {Case} {Study} in the {Transmission} of {Western} {Chant}},
  author={Hornby, Emma},
  year={2002},
  publisher={Ashgate Publishing Ltd.}
}

@article{levy1970italian,
  title={The {I}talian {N}eophytes' {C}hants},
  author={Levy, Kenneth},
  journal={Journal of the American Musicological Society},
  volume={23},
  number={2},
  pages={181--227},
  year={1970},
  publisher={JSTOR}
}

@book{narmour1992analysis,
  title={The analysis and cognition of melodic complexity: The implication-realization model},
  author={Narmour, Eugene},
  year={1992},
  publisher={University of Chicago Press}
}

@article{hajicjr2025genomeOfMelody,
  title = {{Genome of Melody: Applying bioinformatics to study the evolution of Gregorian chant}},
  url = {https://doi.org/10.1098/rstb.2024.0274},
  DOI = {10.1098/rstb.2024.0274},
  journal = {Philosophical Transactions of the Royal Society B: Biological Sciences},
  author = {Hajič jr.,  Jan and Lanz,  Vojtěch and Ballen,  Gustavo A.},
  year = {2025},
  volume = {380},
  number = {1940},
  month = dec 
}

@inproceedings{martinez2023holistic,
  title={A holistic approach for aligned music and lyrics transcription},
  author={Martinez-Sevilla, Juan C and Rios-Vila, Antonio and Castellanos, Francisco J and Calvo-Zaragoza, Jorge},
  booktitle={International conference on document analysis and recognition},
  pages={185--201},
  year={2023},
  organization={Springer}
}

@inproceedings{huglo2009statistics,
    author = {Huglo, Michel},
    title = {Statistical {Survey} of {Notated Liturgical Manuscripts}},
    booktitle = {Antiphonaria: Studien zu Quellen und Gesängen des Mittelalterlichen Offiziums},
    year = {2009},
    volume = {7},
    series = {Regensburger Studien zur Musikgeschichte},
    editor = {Horn, Wolfgang and Hiley, David}
}

@inproceedings{eipert2019editor,
  title={Editor support for digital editions of medieval monophonic music},
  author={Eipert, Tim and Herrmann, Felix and Wick, Christoph and Puppe, Frank and Haug, Andreas},
  booktitle={Proceedings of the 2nd International Workshop on Reading Music Systems},
  pages={4--7},
  year={2019}
}

@inproceedings{wick2019ommr4all,
  title={{OMMR4all} — a {Semiautomatic} {Online} {Editor} for {Medieval} {Music} {Notations}},
  author={Wick, Christoph and Puppe, Frank},
  booktitle={2nd International workshop on reading music systems},
  pages={31--34},
  year={2019}
}

@inproceedings{lanz2025segmentation,
    author = {Lanz, Vojtěch and Hajič jr., Jan},
    title = {Gregorian {Melody}, {Modality}, and {Memory}: Segmenting {Chant} with {Bayesian} {Nonparametrics}},
    booktitle = {Proceedings of the 26th International Society for Music Information Retrieval Conference},
    year = {2025},
    month = sep,
    pages={638--646},
    doi={10.5281/zenodo.17706544},
    url={https://zenodo.org/records/17706545}
}

@inproceedings{lanz2025chantlab,
    author = {Lanz, Vojtěch and Szabová, Kristina and Hajič jr., Jan},
    title = {Making computational study of {Gregorian} melody accessible with {ChantLab}},
    booktitle = {Music Encoding Conference 2025 Book of Abstracts},
    year = {2025},
    month = jun,
    pages={157-163},
    doi={10.17613/z50gm-qf714},
    url= {https://works.hcommons.org/records/z50gm-qf714}
}

@inproceedings{dvorakova2025chantmapper,
    author = {Dvořáková, Anna and Hajič jr., Jan},
    title = {Visualising {G}regorian {T}raditions: Chant{M}apper},
    booktitle = {Music Encoding Conference 2025 Book of Abstracts},
    year = {2025},
    month = jun,
    pages={143-151},
    doi={10.17613/20s0d-gq678},
    url={https://works.hcommons.org/records/g378z-ppk48}
}

@inproceedings{phan2025analyser,
    author = {Phan, Antoine and Thomae, Martha E. and De Luca, Elsa and Oriol, Francesco},
    title = {Analyser {T}ool for {MEI} {N}eumes {E}ncoded {C}hants},
    booktitle = {Music Encoding Conference 2025 Book of Abstracts},
    year = {2025},
    month = jun,
    pages={36-48},
    doi={10.17613/jm6rw-btm49},
    url={https://works.hcommons.org/records/jm6rw-btm49}
}

@inproceedings{hajicjr2025unseen,
    author = {Hajič jr., Jan and Moss, Fabian},
    title = {Knowing when to stop: insights from ecology for building catalogues, collections, and corpora},
    booktitle = {Proceedings of the 12th Digital Libraries for Musicology Conference},
    year = {2025},
    month = sep,
    pages = {90-94},
    doi = {10.1145/3748336.3748347},
    url={https://doi.org/10.1145/3748336.3748347}
}

@article{eipert2025corpus,
  title={Corpus {Troporum} {Dataset}: {A} {Digital} {Catalog} of {Trope} {Elements} in {Medieval} {Chant}},
  author={Eipert, Tim and Bongartz, Corinna and Moss, Fabian C},
  journal={Journal of Open Humanities Data},
  volume={11},
  number={1},
  year={2025}
}

@article{hornby2022CEAP,
  title = {Chant editing and analysis program: a tool for analyzing liturgical chant},
  volume = {14},
  ISSN = {1754-6567},
  url = {http://dx.doi.org/10.1080/17546559.2021.2023752},
  DOI = {10.1080/17546559.2021.2023752},
  number = {1},
  journal = {Journal of Medieval Iberian Studies},
  publisher = {Informa UK Limited},
  author = {Hornby,  Emma and Maloy,  Rebecca and Rouse,  Paul},
  year = {2022},
  month = jan,
  pages = {82–95}
}

@book{hesbert1963CAO,
  author = {Hesbert, René Jean},
  address = {Roma},
  booktitle = {Corpus antiphonalium officii},
  language = {lat},
  organization = {Catholic Church.},
  publisher = {Herder},
  series = {Rerum Ecclesiasticarum documenta. Series maior : Fontes ; 7-12},
  title = {Corpus antiphonalium officii },
  year = {1963--1979},
  volume = {1--6}
}

@article{moss2025rest,
  title = {The Rest is Silence: Leveraging Unseen Species Models for Computational Musicology},
  author = { Moss, Fabian C and Hajič jr., Jan and Nachtwey, Adrian and Pugin, Laurent},
  year = {2025},
  journal = {Anthology of Computers and the Humanities},
  volume = {3},
  pages = {557--574},
  editor = {Taylor Arnold, Margherita Fantoli, and Ruben Ros},
  doi = {10.63744/tP4bLwLkye8B},
  url={https://doi.org/10.63744/tP4bLwLkye8B}
}

@book{hiley2009gregorian,
    author = {David Hiley},
    title = {Gregorian Chant. Cambridge Introductions to Music.},
    publisher = {Cambridge University Press},
    year = {2009}
}

@article{huron2006cognitive,
  title={A {Cognitive} {Approach} to {Medieval} {Mode}: {Evidence} for an {Historical} {Antecedent} to the {Major/Minor} {System}},
  author={David Huron and Joshua Veltman},
  journal={Empirical Musicology Review},
  year={2006},
  volume={1},
  pages={33-55},
  url={https://api.semanticscholar.org/CorpusID:37288543},
  doi = {10.18061/1811/24072}
}

@article{FuentesMartinez2026amnlt,
    title = {Aligned music notation and lyrics transcription},
    journal = {Pattern Recognition},
    volume = {170},
    pages = {112094},
    year = {2026},
    issn = {0031-3203},
    doi = {https://doi.org/10.1016/j.patcog.2025.112094},
    url = {https://www.sciencedirect.com/science/article/pii/S003132032500754X},
    author = {Eliseo Fuentes-Martínez and Antonio Ríos-Vila and Juan C. Martinez-Sevilla and David Rizo and Jorge Calvo-Zaragoza},
}

@article{McBride2025commentary,
  title = {Commentary on {B}uechele et al. (2023): {Communicating} {Across} the {Divide} – a {Place} for {Physics} in {Music}?},
  volume = {19},
  ISSN = {1559-5749},
  url = {http://dx.doi.org/10.18061/emr.v19i2.9783},
  DOI = {10.18061/emr.v19i2.9783},
  number = {2},
  journal = {Empirical Musicology Review},
  publisher = {The Ohio State University Libraries},
  author = {McBride,  John M.},
  year = {2025},
  month = mar,
  pages = {154–172}
}

@incollection{solesmes1960groups,
    author = {Monks of Solesmes, of},
    title = {IV: Le Texte neumatique, i: Le Groupement des manuscrits},
    booktitle = {Graduel romain: Edition critique par les moines de Solesmes},
    publisher = {Abbey of Solesmes},
    year = {1960}
}

%
%
%
%

\end{document}